\documentclass[conference,a4paper]{IEEEtran}
\usepackage{xcolor}

\usepackage{cite}

\ifCLASSINFOpdf
  \usepackage[pdftex]{graphicx}
\else
  \usepackage[dvips]{graphicx}
\fi
\usepackage[cmex10]{amsmath}
\usepackage{amssymb}
\begin{document}
%
\title{Perfect Anomalous Refraction with Metagratings}

\author{\IEEEauthorblockN{
Ariel Epstein* and   
Oshri Rabinovich
}                                     
\IEEEauthorblockA{Andrew and Erna Viterbi Faculty of Electrical Engineering, Technion - Israel Institute of Technology, Haifa 3200003, Israel}
 \IEEEauthorblockA{*epsteina@ee.technion.ac.il}
}



\maketitle

\begin{abstract}
We present a methodology for designing metagratings for perfect anomalous refraction, based on multilayered loaded wire arrays. In recent work, it has been shown that such structures can implement perfect anomalous deflection and beam splitting in \emph{reflect-mode}, using only a handful of subwavelength meta-atoms per (wavelength-scale) macro-period. Extending previous formulations to enable manipulation of \emph{transmitted} fields as well, we derive analytical relations between the scattered fields, currents induced on the wires, and the individual load impedances, and enforce conditions that guarantee elimination of spurious scattering while retaining a passive and lossless structure. Utilizing our recent results, we demonstrate how the multilayered metagratings can be realized using realistic printed-capacitor loads, whose geometry can be analytically resolved. Thus, this design scheme, which can be fully implemented in MATLAB, prescribes simple physical structures, achieving optimal anomalous refraction efficiency without requiring even a single full-wave simulation. This paves the path for harnessing this novel concept for applications requiring control on transmitted diffraction modes (e.g., lenses), taking advantage of an efficient and rigorous design scheme, and simplified structure.
\end{abstract}


%
\IEEEpeerreviewmaketitle

\vspace{7pt}
\section{Introduction}
\label{sec:introduction}
Diffraction gratings have been extensively investigated in the past as means to implement desirable engineered diffraction, for instance, coupling an incoming beam into a certain reflection or transmission mode \cite{Loewen1997}. The fundamental physical principle underlying their operation is related to the Floquet-Bloch (FB) theorem, stating that a periodic planar structure may facilitate coupling of an incident plane wave to a discrete set of propagating and evanescent modes, whose transverse wavenumbers are determined by the structure's period \cite{Russell1986}. Consequently, the basic design methodology followed by most reports attempts to establish efficient coupling to a desirable FB mode, by choosing a period such that only a handful of propagating modes are allowed, and employing some heuristics or numerical optimization of a single period to diminish the coupling to all modes other than the prescribed one.

Although highly-efficient diffraction gratings have been reported in the past \cite{Clausnitzer2008}, a rigorous formulation guaranteeing optimal performance for general scenarios of engineered (anomalous) refraction and reflection was not presented. In recent years, work on electromagnetic metasurfaces have lead to breakthroughs in this area. These ultrathin devices, comprised of closely-packed subwavelength polarizable particles, were shown to be well modelled by equivalent generalized sheet transition conditions (GSTCs), relating the field discontinuity they induce to the metasurface constituents \cite{Kuester2003,Epstein2016_2}. Based on this approach, polarizability distributions that can accurately implement arbitrary (even extreme) anomalous refraction and reflection with unitary coupling efficiencies were rigorously derived, based on omega-bianisotropic metasurfaces \cite{Wong2016,Epstein2016_3,Asadchy2016,Epstein2016_4}. Nonetheless, practical realization of such metasurfaces, featuring three degrees of freedom per meta-atom and numerous meta-atoms per wavelength, revealed to be a nontrivial task, typically relying on time-consuming full-wave simulations \cite{Pfeiffer2014_2, Asadchy2015, Achouri2015_1, Epstein2016}. Consequently, alternative approximate methodologies have been developed based on original GSTC-oriented physical intuition, which reduced the design complexity and facilitated experimental demonstrations \cite{Estakhri2016_1, Estakhri2017, Asadchy2017, DiazRubio2017}.

This quest for simplified realizations of perfect anomalous reflecting and refracting surfaces has led various researchers to revisit the concept of diffraction gratings of late. Inspired by the advances in metasurface research, numerous recent reports explored the possibility to realize these functionalities using periodic planar arrays of one or two subwavelength inclusions (meta-atoms) per macro period, also known as metagratings \cite{Memarian2017,Radi2017,Chalabi2017,Sell2017,Khaidarov2017,Wong2017_2}; similar to diffraction gratings, metagrating operation relies on elimination of undesirable FB modes. 

\begin{figure*}[!t]
\centering
\includegraphics[width=16cm]{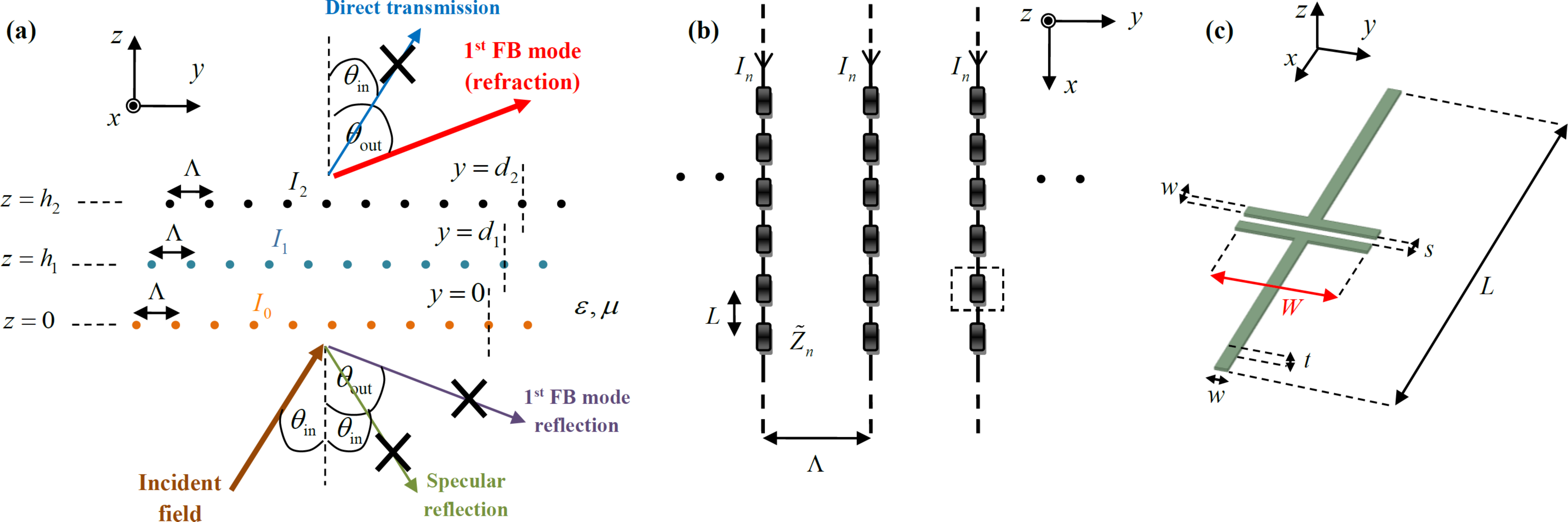}
\caption{Physical configuration of a multilayered loaded-wire metagrating refracting a plane wave incident from $\theta_\mathrm{in}$ towards $\theta_\mathrm{out}$, while eliminating all spurious FB modes. (a) Front view of the complete structure. (b) Top view of a single metagrating layer. (c) Trimetric view of a single unit cell [denoted by a dashed rectangle in (b)], where the load is implemented by a printed capacitor.}
\label{fig:physical_configuration}
\end{figure*}

Maybe the most rigorous treatment of this configuration for perfect anomalous \emph{reflection} was presented in a recent paper by Ra'di \textit{et al.} \cite{Radi2017}, based on the analytical framework developed in \cite{Wait1954,Tretyakov2003}. In this work the authors analyzed a periodic array of identical magnetically-polarizable scatterers (conducting loops) with adjustable loads, placed in front of a perfect electric conductor (PEC) plane, where the periodicity was chosen as to allow only a single propagating FB mode other than the fundamental. Subsequently, by formulating the conditions on the currents induced on the meta-atom array that would guarantee elimination of specular reflection and conservation of real power, they reached an equation for the array-PEC separation distance required to ensure full coupling of the incoming plane wave to the desirable reflected mode. Once this distance was resolved, a short sweep of the meta-atom load in a commercial full-wave solver finalized the design procedure. The resultant simple structure, based on a repeating single subwavelength element, demonstrated perfect anomalous reflection, even at extreme angles. A similar approach was applied by the same group to reflecting metagratings made of dielectric cylinders above a PEC, now using two meta-atoms per macro period \cite{Chalabi2017}; the additional degree of freedom, i.e. the second cylinder, enabled elimination of two spurious FB modes, when required (this was also demonstrated in \cite{Radi2017} by using a single bianisotropic meta-atom, and in \cite{Wong2017_2} where the additional scatterer was needed due to the prefixed value of the array-PEC separation distance).

More recently, we have presented a detailed analytical model for reflecting metagratings based on loaded conducting \emph{wires} in front of a PEC \cite{Epstein2017_1}. In contrast to \cite{Radi2017, Chalabi2017}, such a configuration is compatible with printed-circuit-board (PCB) fabrication, commonly used for realizing devices at microwave frequencies. Therein, we have shown that strips loaded by printed capacitors can implement perfect reflect-mode beam splitting of a transverse electric (TE) polarized normally-incident plane wave towards $\pm\theta_\mathrm{out}$. More than that, by further developing the analytical formulation to relate the individual loads and the incident field \cite{Tretyakov2003}, we have devised a semianalytical design methodology that prescribes the required geometry of the copper traces to achieve optimal coupling. The same analytical framework was used to analyze the sensitivity of the synthesized metagrating beam splitters to conductor losses, fabrication inaccuracies, and frequency shifts \cite{Epstein2017_1}. The translation of such loaded-strip reflective metagratings into a fabrication-ready PCB device can also be achieved analytically, as we have shown in \cite{Rabinovich2017} by rigorously incorporating the dielectric substrate into the analytical formalism.

In this paper, we present a methodology to realize metagratings for anomalous \emph{refraction}, i.e. for coupling a plane wave incoming from $\theta_\mathrm{in}$ into a designated \emph{transmission} angle $\theta_\mathrm{out}$. This is achieved by using multilayered arrays of capacitively-loaded wires, which provides the sufficient degrees of freedom to control the increased number of possible FB modes. Modifying the analytical model of \cite{Epstein2017_1,Rabinovich2017} to include cascaded loaded strips as in \cite{Ikonen2007}, we formulate the necessary conditions on the capacitive loads to eliminate all undesirable FB modes \emph{and} guarantee passive and lossless design specifications. Resolving these conditions using a simple MATLAB program, we show that perfect refracting metagrating designs can be specified up to the conductor trace geometry, without conducting \emph{even a single} full-wave simulation. This methodology extends the range of applications of these simple devices for advanced diffraction engineering also to transmit-mode devices, while retaining the appealing analytical rigour and efficiency, and reduced realization complexity.

\vspace{7pt}
\section{Theory}
\label{sec:theory}
We consider a 2D ($\partial/\partial x=0$) $\Lambda$-periodic configuration excited by TE ($E_z=E_y=H_x=0$) polarized fields, in which three capacitively-loaded wires are situated at $\left(y,z\right)=\left(0,0\right)$, $\left(y,z\right)=\left(d_1,h_1\right)$, and $\left(y,z\right)=\left(d_2,h_2\right)$ [Fig. \ref{fig:physical_configuration}(a)]. The wire-arrays are embedded in a medium with permittivity $\epsilon$ and permeability $\mu$, defining the wavenumber $k=\omega\sqrt{\mu\epsilon}$ and the wave impedance $\eta=\sqrt{\mu/\epsilon}$; harmonic time dependency $e^{j\omega t}$ is assumed and suppressed. The wires are loaded by a distributed impedance of $\tilde{Z}_0$, $\tilde{Z}_1$, and $\tilde{Z}_2$, respectively, and we denote the corresponding currents as $I_0$, $I_1$, and $I_2$ [Fig. \ref{fig:physical_configuration}(b)].

The structure is excited by a plane wave with an angle of incidence $\theta_\mathrm{in}$, and our goal is to couple all this incoming power to the refracted plane wave, departing at an angle of $\theta_\mathrm{out}$; according to the FB theorem \cite{Tretyakov2003}, this dictates that $\Lambda=\lambda/\left|\sin\theta_\mathrm{out}-\sin\theta_\mathrm{in}\right|$, where $\lambda=2\pi/k$ is the operating wavelength. To reduce as much as possible the number of propagating FB modes that we need to control, we restrict ourselves to the cases in which \cite{Radi2017,Rabinovich2017}
\begin{equation}
\!\begin{array}{l}
\vspace{2pt}
\theta_\mathrm{out}\!\!\in\!\!\left(-\frac{\pi}{2},\arcsin\!\left\{\!2\sin\theta_\mathrm{in}-1\!\right\}\right) \wedge \theta_\mathrm{in}\!\!\in\!\!\left(0,\arcsin\!\left\{\!\frac{1}{3}\!\right\}\right) \\
\vspace{2pt}
\,\,\,\,\,\mathrm{or} \\
\vspace{2pt}
\theta_\mathrm{out}\!\!\in\!\!\left(-\frac{\pi}{2},\arcsin\!\left\{\!\frac{1}{2}\sin\theta_\mathrm{in}-\frac{1}{2}\!\right\}\right) \wedge \theta_\mathrm{in}\!\!\in\!\!\left(\arcsin\!\left\{\!\frac{1}{3}\!\right\}\!\!,\frac{\pi}{2}\right)
\end{array}\!\!\!
\label{equ:period}
\end{equation}
where we assume, without loss of generality, that $\theta_\mathrm{in}>0$. For angles of incidence and departure that satisfy \eqref{equ:period}, only the fundamental and the first FB modes (in transmission and reflection) are propagating. Therefore, to guarantee full coupling of the incident plane wave to the first mode in transmission (propagating towards $\theta_\mathrm{out}$) we need to eliminate coupling to the other three propagating FB modes [see Fig. \ref{fig:physical_configuration}(a)]: specular reflection (green arrow), direct transmission (blue arrow), and the first mode in reflection (purple arrow).

\begin{figure*}[!t]
\normalsize
\setcounter{equation}{1}
\begin{equation}
\begin{array}{l}
	\vspace{2pt}	
	E_x^\mathrm{tot}\left(y,z\right)=E_\mathrm{in}e^{-jky\sin\theta_\mathrm{in}}e^{-jkz\cos\theta_\mathrm{in}} \\
	 -\dfrac{I_0}{2\Lambda}\!\!\left[\!Z_\mathrm{in}e^{-jky\sin\theta_\mathrm{in}}e^{-jk\left|z\right|\cos\theta_\mathrm{in}}
		\!\!+\!\!Z_\mathrm{out}e^{-jky\sin\theta_\mathrm{out}}e^{-jk\left|z\right|\cos\theta_\mathrm{out}}
		\!\!-\!\!jk\eta\!\!\!\!\!\!\!\!
		\displaystyle\sum\limits_{\scriptsize\begin{array}{l}m\!\!=\!\!-\infty\\m\!\!\neq\!0,1\end{array}}^{\infty}
			\!\!\!\!\!\!\!e^{-jk_{t,m}y}\dfrac{e^{-\alpha_m\left|z\right|}}{\alpha_m}
	\!\right] \\ 	\vspace{2pt}	
	 -\dfrac{I_1}{2\Lambda}\!\!\left[\!Z_\mathrm{in}e^{-jk\left(y-d_1\right)\sin\theta_\mathrm{in}}e^{-jk\left|z-h_1\right|\cos\theta_\mathrm{in}}
		\!\!+\!\!Z_\mathrm{out}e^{-jk\left(y-d_1\right)\sin\theta_\mathrm{out}}e^{-jk\left|z-h_1\right|\cos\theta_\mathrm{out}}
		\!\!-\!\!jk\eta\!\!\!\!\!\!\!\!
		\displaystyle\sum\limits_{\scriptsize\begin{array}{l}m\!\!=\!\!-\infty\\m\!\!\neq\!0,1\end{array}}^{\infty}
			\!\!\!\!\!\!\!e^{-jk_{t,m}\left(y-d_1\right)}\dfrac{e^{-\alpha_m\left|z-h_1\right|}}{\alpha_m}
	\!\right]\!\!
	\\ 	\vspace{2pt}	
	 -\dfrac{I_2}{2\Lambda}\!\!\left[\!Z_\mathrm{in}e^{-jk\left(y-d_2\right)\sin\theta_\mathrm{in}}e^{-jk\left|z-h_2\right|\cos\theta_\mathrm{in}}
		\!\!+\!\!Z_\mathrm{out}e^{-jk\left(y-d_2\right)\sin\theta_\mathrm{out}}e^{-jk\left|z-h_2\right|\cos\theta_\mathrm{out}}
		\!\!-\!\!jk\eta\!\!\!\!\!\!\!\!
		\displaystyle\sum\limits_{\scriptsize\begin{array}{l}m\!\!=\!\!-\infty\\m\!\!\neq\!0,1\end{array}}^{\infty}
			\!\!\!\!\!\!\!e^{-jk_{t,m}\left(y-d_2\right)}\dfrac{e^{-\alpha_m\left|z-h_2\right|}}{\alpha_m}
	\!\right]\!\!
\end{array}
\label{equ:fields_above_below}
\end{equation}
\hrulefill
\setcounter{equation}{0}
\end{figure*}
\setcounter{equation}{2}

The fields 
everywhere (excluding the wire positions) can be formulated as a superposition of the incident plane wave (external field) and the sum of fields generated by the individual current-carrying wires. The latter correspond to three infinite sums of shifted Hankel functions, which can be transformed into three rapidly-converging series of FB modes via the Poisson formula \cite{Tretyakov2003,Ikonen2007,Epstein2017_1,Rabinovich2017}. Consequently, for a given incident field amplitude $E_\mathrm{in}$, the total fields at any observation point (away from the wires) are given by \eqref{equ:fields_above_below} at the top of this page, where $Z_\mathrm{in}=\eta/\cos\theta_\mathrm{in}$ and $Z_\mathrm{out}=\eta/\cos\theta_\mathrm{out}$ are, respectively, the wave impedances of the incident and refracted plane waves, the transverse wavenumber of the $m$th FB mode is $k_{t,m}=k\sin\theta_\mathrm{in}+mk\left(\sin\theta_\mathrm{out}-\sin\theta_\mathrm{in}\right)$, and the decay coefficient of the $m$th mode ($m\neq0,1$) is $\alpha_m=\left(k_{t,m}^2-k^2\right)^{1/2}>0$.

Rearranging the terms in \eqref{equ:fields_above_below} according to their modal contribution, we can identify the coefficients responsible for excitation of specular reflection (terms proportional to $e^{jkz\cos\theta_\mathrm{in}}$ for $z<0$), reflection of the first mode (proportional to $e^{jkz\cos\theta_\mathrm{out}}$ for $z<0$), and direct transmission (proportional to $e^{-jkz\cos\theta_\mathrm{in}}$ for $z>h_2$). To ensure that no power is carried to the far field via these spurious modes, the metagrating has to be designed such that all these three coefficients vanish. This yields the following conditions on the induced currents
\begin{equation}
\left\{\!\!\!\!
\begin{array}{l}
\vspace{2pt}
I_0\!\!+\!\!I_1 e^{jk(d_1\!\sin\theta_\mathrm{in}-h_1\!\cos\theta_\mathrm{in})}\!\!+\!\!I_2e^{jk(d_2\!\sin\theta_\mathrm{in}-h_2\!\cos\theta_\mathrm{in})}\!\!=\!\!0 \\
\vspace{2pt}
I_0\!\!+\!\!I_1 e^{jk(d_1\!\sin\theta_\mathrm{out}-h_1\!\cos\theta_\mathrm{out})}\!\!+\!\!I_2e^{jk(d_2\!\sin\theta_\mathrm{out}-h_2\!\cos\theta_\mathrm{out})}\!\!=\!\!0 \\
\vspace{2pt}
I_0\!\!+\!\!I_1 e^{jk(d_1\!\sin\theta_\mathrm{in}+h_1\!\cos\theta_\mathrm{in})}\!\!+\!\!I_2e^{jk(d_2\!\sin\theta_\mathrm{in}+h_2\!\cos\theta_\mathrm{in})}\!\!=\!\!\frac{2\Lambda}{Z_\mathrm{in}}E_\mathrm{in}
\end{array}
\right.
\label{equ:eliminating_spurious_modes}
\end{equation}
These can be treated as a set of three linear equations with three unknowns $I_0$, $I_1$, and $I_2$, which can be resolved to yield explicit expressions for the required induced currents as a function of the given incident field amplitude $E_\mathrm{in}$ and the yet-to-be-determined wire offsets $d_1,d_2,h_1,h_2$. For brevity, we do not include herein the closed-form expressions, but rather denote these solutions as $I_n=E_\mathrm{in}\xi_n\left(d_1,d_2,h_1,h_2\right)$, for ${n=0,1,2}$, with the functions $\xi_n$ given by \eqref{equ:eliminating_spurious_modes}.

Next, we derive the relations between the individual loads and the metagrating configuration that would lead to a self-consistent excitation of the currents prescribed by \eqref{equ:eliminating_spurious_modes}. To this end, we use Ohm's law in conjunction with the total fields applied \emph{on} the wires \cite{Tretyakov2003, Ikonen2007, Epstein2017_1, Rabinovich2017}. For a given wire, these are composed of the external fields at the wire position, the mutual fields impressed on the wire due to currents flowing on the other wires (both within the same layer and on the other layers of the metagrating), and the self-induced fields by the currents flowing on the wire itself. The latter require consideration of the wire finite physical dimensions, due to their divergence at the center of the wire; more specifically, the self-fields are evaluated on the wire outer shell, given by an effective radius $r_\mathrm{eff}=w/4$ \cite{Tretyakov2003}, where $w$ is the strip width [Fig. \ref{fig:physical_configuration}(c)].

Overall, the fields on the reference wires at $\left(y,z\right)=\left(0,0\right)$, $\left(y,z\right)=\left(d_1,h_1\right)$, and $\left(y,z\right)=\left(d_2,h_2\right)$, are given, respectively, by \eqref{equ:fields_at_wires} at the top of the next page, where the Hankel functions can be transformed into series with better convergence properties as demonstrated in \cite{Tretyakov2003, Ikonen2007, Epstein2017_1}. Substituting the expressions $I_n=E_\mathrm{in}\xi_n\left(d_1,d_2,h_1,h_2\right)$, retrieved from \eqref{equ:eliminating_spurious_modes}, into \eqref{equ:fields_at_wires}, allows us to establish a direct relation between the required distributed impedances and the wire positions, namely, $\tilde{Z}_n=\zeta_n\left(d_1,d_2,h_1,h_2\right)$, for $n=0,1,2$.

\begin{figure*}[!t]
\normalsize
\setcounter{equation}{3}
\begin{equation}
\begin{array}{l}
	E_x\left(y=0,z=0\right)= \tilde{Z}_0I_0 = E_\mathrm{in}-\frac{k\eta}{4}I_0 H_0^{\left(2\right)}\left(kr_\mathrm{eff}\right)
		-\frac{k\eta}{4}I_0\!\!\!\!\!\!
		\displaystyle\sum\limits_{\scriptsize\begin{array}{l}n\!\!=\!\!-\infty\\n\!\!\neq\!0\end{array}}^{\infty}
		\!\!\!\!\!e^{-jkn\Lambda\sin\theta_\mathrm{in}}H_0^{\left(2\right)}\left(k\left|n\Lambda\right|\right) \\ \vspace{2pt}
	\,\,\,\,\,\,\,\, -\frac{k\eta}{4}I_1\!\!\!\!
		\displaystyle\sum\limits_{n=-\infty}^{\infty}
		\!\!\!e^{-jkn\Lambda\sin\theta_\mathrm{in}}H_0^{\left(2\right)}\left(k\sqrt{\left(d_1+n\Lambda\right)^2+h_1^2}\right)	
		-\frac{k\eta}{4}I_2\!\!\!\!
		\displaystyle\sum\limits_{n=-\infty}^{\infty}
		\!\!\!e^{-jkn\Lambda\sin\theta_\mathrm{in}}H_0^{\left(2\right)}\left(k\sqrt{\left(d_2+n\Lambda\right)^2+h_2^2}\right)  \\	
		
	E_x\left(y=d_1,z=h_1\right)=\tilde{Z}_1I_1=E_\mathrm{in}e^{-jkd_1\sin\theta_\mathrm{in}}e^{-jkh_1\cos\theta_\mathrm{in}}-\frac{k\eta}{4}I_1 H_0^{\left(2\right)}\left(kr_\mathrm{eff}\right)
		-\frac{k\eta}{4}I_1\!\!\!\!\!\!
		\displaystyle\sum\limits_{\scriptsize\begin{array}{l}n\!\!=\!\!-\infty\\n\!\!\neq\!0\end{array}}^{\infty}
		\!\!\!\!\!e^{-jkn\Lambda\sin\theta_\mathrm{in}}H_0^{\left(2\right)}\left(k\left|n\Lambda\right|\right) \\
		\vspace{2pt}
	\,\,\,\,\,\,\,\, -\frac{k\eta}{4}I_0\!\!\!\!
		\displaystyle\sum\limits_{n=-\infty}^{\infty}
		\!\!\!e^{-jkn\Lambda\sin\theta_\mathrm{in}}H_0^{\left(2\right)}\left(k\sqrt{\left(d_1-n\Lambda\right)^2+h_1^2}\right)	
		-\textstyle\frac{k\eta}{4}I_2\!\!\!\!
		\displaystyle\sum\limits_{n=-\infty}^{\infty}
		\!\!\!
		e^{-jkn\Lambda\sin\theta_\mathrm{in}}H_0^{\left(2\right)}\left(k\sqrt{\left(d_2-d_1+n\Lambda\right)^2+\left(h_2-h_1\right)^2}\right)	\\	
		
	E_x\left(y=d_2,z=h_2\right)=\tilde{Z}_2I_2=E_\mathrm{in}e^{-jkd_2\sin\theta_\mathrm{in}}e^{-jkh_2\cos\theta_\mathrm{in}}-\frac{k\eta}{4}I_2 H_0^{\left(2\right)}\left(kr_\mathrm{eff}\right)
		-\frac{k\eta}{4}I_2\!\!\!\!\!\!
		\displaystyle\sum\limits_{\scriptsize\begin{array}{l}n\!\!=\!\!-\infty\\n\!\!\neq\!0\end{array}}^{\infty}
		\!\!\!\!\!e^{-jkn\Lambda\sin\theta_\mathrm{in}}H_0^{\left(2\right)}\left(k\left|n\Lambda\right|\right) \\
		\vspace{2pt}
	\,\,\,\,\,\,\,\, -\frac{k\eta}{4}I_0\!\!\!\!
		\displaystyle\sum\limits_{n=-\infty}^{\infty}
		\!\!\!e^{-jkn\Lambda\sin\theta_\mathrm{in}}H_0^{\left(2\right)}\left(k\sqrt{\left(d_2-n\Lambda\right)^2+h_2^2}\right)	
		-\textstyle\frac{k\eta}{4}I_1\!\!\!\!
		\displaystyle\sum\limits_{n=-\infty}^{\infty}
		\!\!\!
		e^{-jkn\Lambda\sin\theta_\mathrm{in}}H_0^{\left(2\right)}\left(k\sqrt{\left(d_2-d_1-n\Lambda\right)^2+\left(h_2-h_1\right)^2}\right)		
\end{array}
\label{equ:fields_at_wires}
\end{equation}
\hrulefill
\setcounter{equation}{0}
\end{figure*}
\setcounter{equation}{4}

At this point, for any given wire positions $\left(d_1,d_2,h_1,h_2\right)$, setting the load impedances in the metagrating of Fig. \ref{fig:physical_configuration} following $\tilde{Z}_n=\zeta_n\left(d_1,d_2,h_1,h_2\right)$ would yield induced currents as prescribed in \eqref{equ:eliminating_spurious_modes}, i.e., $I_n=E_\mathrm{in}\xi_n\left(d_1,d_2,h_1,h_2\right)$; thus, such load impedances would guarantee that no spurious diffraction orders would be scattered from the structure. However, we have yet to enforce any condition on power conservation within the structure; therefore, for arbitrary $\left(d_1,d_2,h_1,h_2\right)$ each of these load impedances could exhibit gain or loss, and the final power radiated towards $\theta_\mathrm{out}$ could be larger or smaller than the power at the incoming beam.

As this is clearly undesirable, 
the last step in our synthesis scheme involves determination of the remaining degrees of freedom $\left(d_1,d_2,h_1,h_2\right)$ to ensure that the design specifications would yield a passive and lossless device. To this end, we demand that the load impedances, which are the only source for gain or loss in the system, would be purely reactive. Using the outcome of \eqref{equ:fields_at_wires}, this can be formulated as
\begin{equation}
\begin{array}{l}
\vspace{2pt}
\Re\left\{\zeta_0\left(d_1,d_2,h_1,h_2\right)\right\}=\Re\left\{\zeta_1\left(d_1,d_2,h_1,h_2\right)\right\} \\
\,\,=\Re\left\{\zeta_2\left(d_1,d_2,h_1,h_2\right)\right\}=0
\end{array}
\label{equ:passive_lossless_condition}
\end{equation}
Equation \eqref{equ:passive_lossless_condition} corresponds to three nonlinear equations, forming three nonlinear constraints on $\left(d_1,d_2,h_1,h_2\right)$. These constraints can be readily resolved by any commercial scientific programming environment (e.g., MATLAB), yielding the required horizontal and vertical offsets of the three metagratings with respect to one another to guarantee passive and lossless design. Substituting these offsets into the relations $\tilde{Z}_n=\zeta_n\left(d_1,d_2,h_1,h_2\right)$ extracted after \eqref{equ:fields_at_wires} finalizes the rigorous design procedure, indicating both the multilayer metagrating geometry and the load impedances required to implement the desirable perfect anomalous refraction.

\vspace{7pt}
\section{Results and Discussion}
\label{sec:results}
To verify our synthesis scheme, we design a multilayered metagrating to fully-couple a plane wave incident at $\theta_\mathrm{in}=10^\circ$ to a refracted plane wave departing at $\theta_\mathrm{out}=-70^\circ$, which satisfies \eqref{equ:period}. As outlined in Section \ref{sec:theory}, we utilize \eqref{equ:eliminating_spurious_modes} and \eqref{equ:fields_at_wires} with the given $\theta_\mathrm{in}$ and $\theta_\mathrm{out}$ to define the suitable set of nonlinear constraints \eqref{equ:passive_lossless_condition}; the operating frequency is $f=20\mathrm{GHz}$, and we consider trace widths of $w=3\mathrm{mil}$ [Fig. \ref{fig:physical_configuration}(c)], compatible with typical PCB fabrication capabilities \cite{Epstein2016}. Subsequently, we sweep in MATLAB the values of the unknowns $\left(d_1,d_2,h_1,h_2\right)$ within the ranges $\left(d_1,d_2\right)\in\left[0,\Lambda\right)\times\left[0,\Lambda\right)$ and $\left(h_1,h_2\right)\in\left(0,\lambda\right]\times\left(0,\lambda\right]$, keeping $h_2>h_1$, and find the combination of metagrating offsets that minimizes all three constraints \eqref{equ:passive_lossless_condition}; of course, more sophisticated root-finding numerical procedures can be used instead.

The optimal wire constellation [Fig. \ref{fig:physical_configuration}(a)] was found to be $d_1=0.844\lambda$, $d_2=0.826\lambda$, $h_1=0.150\lambda$, and $h_2=0.409\lambda$. Substituting these values into \eqref{equ:fields_at_wires} and the associated $\tilde{Z}_n=\zeta_n\left(d_1,d_2,h_1,h_2\right)$ yields the required distributed load impedances $\tilde{Z}_0=-j5.19[\eta/\lambda]$, $\tilde{Z}_1=-j4.96[\eta/\lambda]$, and $\tilde{Z}_2=-j6.76[\eta/\lambda]$, which are purely reactive, as expected from \eqref{equ:passive_lossless_condition}. To realize these capacitive loads in practice, we follow the methodology presented in \cite{Epstein2017_1}, utilizing the printed capacitors of Fig. \ref{fig:physical_configuration}(c), with subwavelength repetition along the $x$ axis to emulate the required homogenized distributed impedance. The capacitors are formed by suitable copper trace geometry, with trace width and separation $w=s=3\mathrm{mil}$, thickness $t=18\mathrm{\mu m}$, $x$-periodicity $L=\lambda/10$, and capacitor width $W$. In \cite{Epstein2017_1} we have shown that for such a geometry, the capacitor width required to implement a given distributed impedance $\tilde{Z}_n$ can be approximated by $W_n=2.85K_\mathrm{corr}C_n[\mathrm{mil/fF}]$, where the capacitance is given by $C_n=-1/(2\pi f L\Im\{\tilde{Z}_n\})$; the correction factor at $f=20\mathrm{GHz}$ was evaluated therein as $K_\mathrm{corr}=0.89$. Correspondingly, we finalize the metagrating design by setting the capacitor widths by using these analytical formulas with the evaluated $\tilde{Z}_n$, yielding $W_0=103.0\mathrm{mil}$, $W_1=107.6\mathrm{mil}$, and $W_2=79.1\mathrm{mil}$.

\begin{figure}[htb]
\centering
\includegraphics[width=6cm]{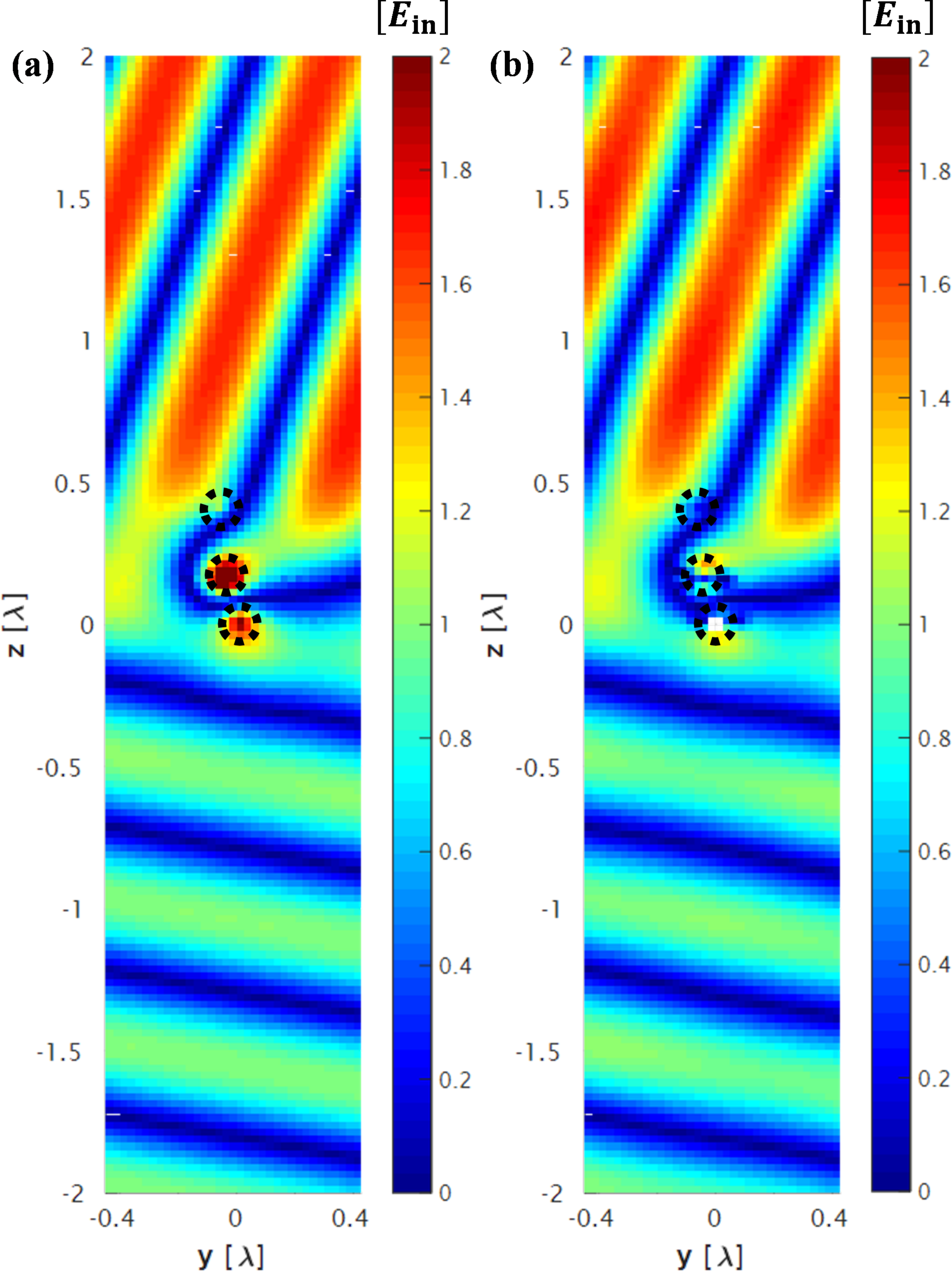}
\caption{Field distribution $\Re\{E_x\left(y,z\right)\}$ over a single period (a) as obtained from the analytical model \eqref{equ:fields_above_below} and (b) as recorded in full-wave simulations, of the designed multilayered metagrating, refracting the incident beam at $\theta_\mathrm{in}=10^\circ$ towards $\theta_\mathrm{out}=-70^\circ$. The regions around the loaded wires, where the predictions are expected to deviate due to the finite size of the printed capacitors, are denoted in black dotted ($\lambda/8$-diameter) circles \cite{Epstein2017_1}.}
\label{fig:results}
\end{figure}

We have defined the prescribed multilayered metagrating in ANSYS HFSS, using copper traces with realistic conductivity of $\sigma=58\times10^6[\mathrm{S/m}]$ for the wires and printed capacitors, and vacuum as the surrounding medium. Without any optimization, full-wave simulation of this periodic structure (using Floquet ports) indicated that $95.7\%$ of the incident power from the incoming plane wave at $\theta_\mathrm{in}=10^\circ$ is coupled to the desirable refracted mode, propagating towards $\theta_\mathrm{out}=-70^\circ$, limited only by inevitable conductor losses, responsible for absorption of $3.9\%$ of the power. As little as $0.4\%$ of the incident power is scattered to spurious FB modes, verifying that indeed, the design procedure leads to highly-efficient perfectly refracting metagrating. The excellent agreement between the analytically-predicted field distribution [Fig. \ref{fig:results}(a)], evaluated using \eqref{equ:fields_above_below}, and the one recorded in full-wave simulations [Fig. \ref{fig:results}(b)], serves as further evidence for the accuracy and validity of the analytical model and synthesis scheme.

\vspace{7pt}
\section{Conclusion}
\label{sec:conclusion}
To conclude, we have formulated a design procedure for perfectly refracting multilayered metagratings, based on a cascade of three capacitively-loaded conducting strips in a unit cell. We have shown that the presented rigorous analytical model can accurately determine the required metagrating offsets with respect to one another, as well as the required copper geometries, without requiring any full-wave simulations. This efficient scheme extends the range of applicability of these simple devices, comprised of only a handful of subwavelength meta-atoms per macro-period (which is of the order of a wavelength), facilitating their usage not only for reflect-mode devices, but also for applications requiring transmit-mode functionalities. Most importantly, the ability to accurately set these multiple degrees of freedom without resorting to full-wave solvers forms a distinguishable advantage with respect to both metasurfaces as well as traditional diffraction gratings, expected to enable more sophisticated wavefront manipulations in the future. 


%
%



\bibliographystyle{IEEEtran}
\bibliography{refractingMetagratings}
%
%
%
%

\end{document}